\documentclass{aa}  
\usepackage{graphicx}
\usepackage{txfonts}
\begin{document}
   \title{Photometry of the five marginally studied
open clusters Collinder~74, Berkeley~27, Haffner~8,
NGC~2509 and VdB-Hagen~4}

   \author{Giovanni Carraro
          \inst{1,2}
          \and
           Edgardo Costa\inst{1}
          }

   \offprints{G. Carraro}

\institute{
             $^1$Departamento de Astronom\'ia, Universidad de Chile,
                 Casilla 36-D, Santiago, Chile\\
              $^2$Dipartimento di Astronomia, Universit\`a di Padova,
                 Vicolo Osservatorio 2, I-35122 Padova, Italy\\
                 }

   \date{Received ....; accepted...}

  \abstract
   { The stellar populations in the outer Galactic disk are nowadays a
   subject of wide interest. To contribute to a better picture of
   this part of the Galaxy,  we have studied the nature of five
   marginally    investigated star clusters
   (Collinder~74, Berkeley~27, Haffner~8, NGC~2509, and VdB-Hagen4)
   by means of accurate CCD photometry in the V and I
   pass-bands. These clusters are in fact located in the Third Galactic Quadrant.}
   {We aim to obtain the basic parameters of these objects, which in some
   cases are still disputed in the literature. In the case of VdB-Hagen~4
   we provide the first estimate of its fundamental parameters, while
  for  Haffner~8 we present the first CCD photometry. }
   {The analysis is based on the comparison between field stars decontaminated
   Color Magnitude Diagrams and stellar models. Particular care is devoted to
   the the assessment of the data quality, and the statistical field stars 
   decontamination. The library of stellar isochrones from Girardi et al. (2000)
   is adopted in this study.}
   {The analysis we carried out allowed us to solve a few inconsistencies in the
   literature regarding Haffner~8 and NGC~2509. 
   Collinder~74 is found to be significantly
   older than reported before. VdB-Hagen~4 is a young open cluster located more than
    20 kpc from the Galactic center. Such an extreme distance is compatible
   with the cluster belonging to the Norma-Cygnus  arm.}
   {}

   \keywords{Open clusters and associations: general- Open clusters
             and associations: individual: Collinder~74, Berkeley~27,
             Haffner~8, NGC~2509, VdB-Hagen4  }

   \maketitle
%

\section{Introduction}
The main reason to acquire and analyze photometry of star
clusters in the Galactic disk resides in the possibility
to build up their Color Magnitude Diagram (CMD), which
allows us to infer estimates of the cluster age, distance
and reddening. In turn, these parameters are routinely used
to trace the large scale structure and evolution of the Galactic disk.
In particular the outer Galactic
disk is nowadays a topic of wide interest, with the discovery
of new spiral features (Moitinho et al. 2006a), and the 
intensively studied structure of the outer disk, where the warp and flare
seem to play a major role (Momany et al. 2006), but where
at the same time external objects in the act of engulfing the Milky Way
like the Monoceros Stream (Newberg et al. 2002) 
are claimed to have been detected.\\
In this context stars clusters are well recognized to be valuable
tracers of the disk structure, in a way that robust
estimates of their fundamental parameters is a need to proceed
in our understanding of the details of the Galaxy structure.\\
In an attempt to contribute to all this, we are presenting here
CCD photometry of 5 Galactic fields, which contain the star clusters
Collinder~74, Berkeley~27, Haffner~8, NGC~2509 and VdB-Hagen~4.
These clusters are  all located in the Third Galactic Quadrant,
spanning 50 degrees in longitude and 15 in latitude (see Table~1).\\
\noindent
Some of these clusters (Collinder~74, Berkeley~27, Haffner~18 and NGC~2509)
have already been investigated in the past at different levels,
and for them we are going to provide a close comparison of the data
and the derived results, since in the literature there still exist
some discrepancy in their basic parameters.\\
On the other hand, VdB-Hagen~4 was never studied before. Still,
it seems a very promising target, since it is located
(see Table~1) very close to the center of the 
lively discussed  Canis Major
over-density (Momany et al. 2006).\\

\noindent
The paper is organized as follows.
In Sect~2 we detail the observations and reduction
of the material which is the base of the present study.
Sect~3 illustrates the method adopted to estimate
the radius of each cluster. In Sect.~4 we deal
with field star decontamination before deriving
in Section~5 the clusters basic parameters.
Sect.~6 summarizes the findings of this paper
suggesting some directions for future
investigation.

\begin{table}
\caption{Coordinates of the clusters under investigation
for J2000.0 equinox and have been
visually verified by us.}
\begin{tabular}{ccccc}
\hline
\hline
\multicolumn{1}{c}{Name} &
\multicolumn{1}{c}{$RA$}  &
\multicolumn{1}{c}{$DEC$}  &
\multicolumn{1}{c}{$l$} &
\multicolumn{1}{c}{$b$} \\
\hline
& {\rm $hh:mm:ss$} & {\rm $^{o}$~:~$^{\prime}$~:~$^{\prime\prime}$} & [deg] & [deg] \\
\hline
Collinder 74    & 05:48:30 & +07:24:00 & 198.97 & -10.41 \\
Berkeley  27    & 06:51:18 & +05:46:00 & 207.78 & +02.60 \\
Haffner 8       & 07:23:24 & -12:20:00 & 227.53 & +01.34 \\
NGC 2509        & 08:00:48 & -19:03:06 & 237.85 & +05.82 \\
VdB-Hagen 4     & 07:37:44 & -36:04:00 & 249.96 & -07.10 \\
\hline\hline
\end{tabular}
\end{table}
 
\

   \begin{figure}
   \centering
   \includegraphics[width=\columnwidth]{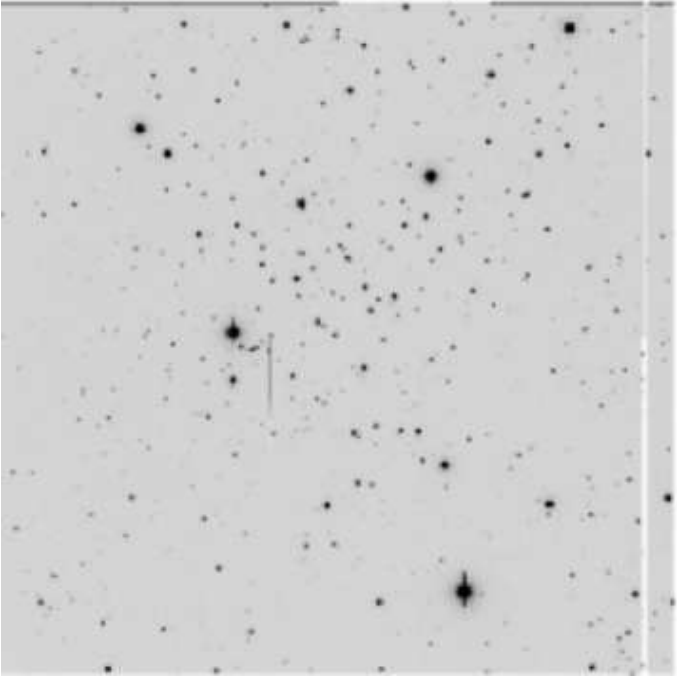}
   \caption{I = 600 sec image of Collinder~74. The field
is 13.5 arcmin on a side. North  is up, East to the left.}%
    \end{figure}

\noindent

   \begin{figure}
   \centering
   \includegraphics[width=\columnwidth]{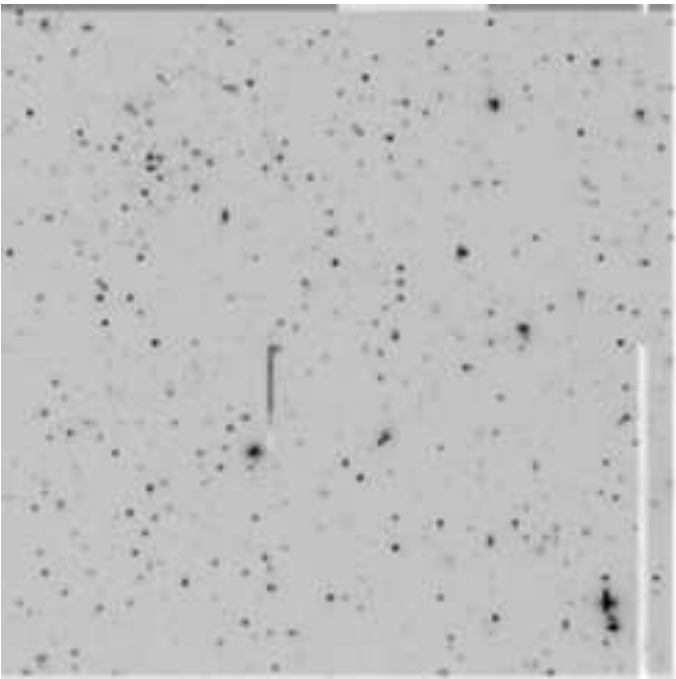}
   \caption{I = 600 sec image of Berkeley~27. The field
is 13.5 arcmin on a side. North  is up, East to the left.}%
    \end{figure}

 \begin{figure}
   \centering
   \includegraphics[width=\columnwidth]{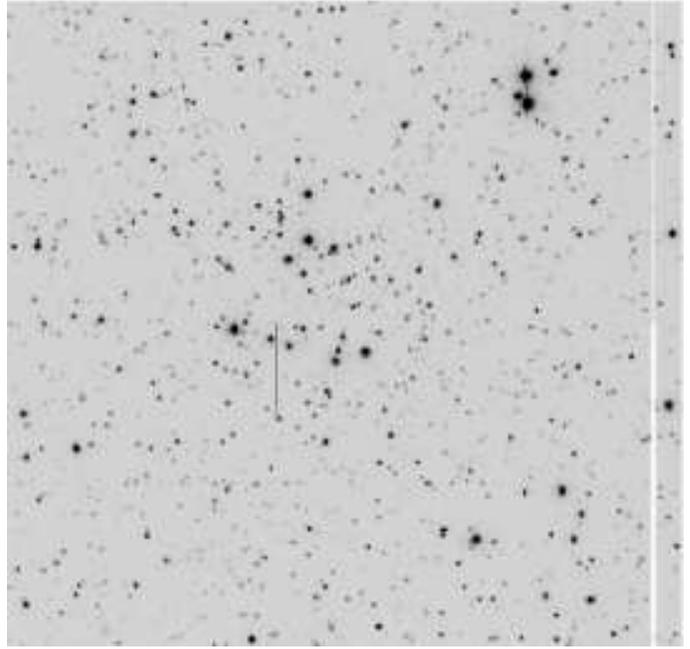}
   \caption{I = 600 sec image of Haffner~8. The field
is 13.5 arcmin on a side. North  is up, East to the left.}%
    \end{figure}

   \begin{figure}
   \centering
   \includegraphics[width=\columnwidth]{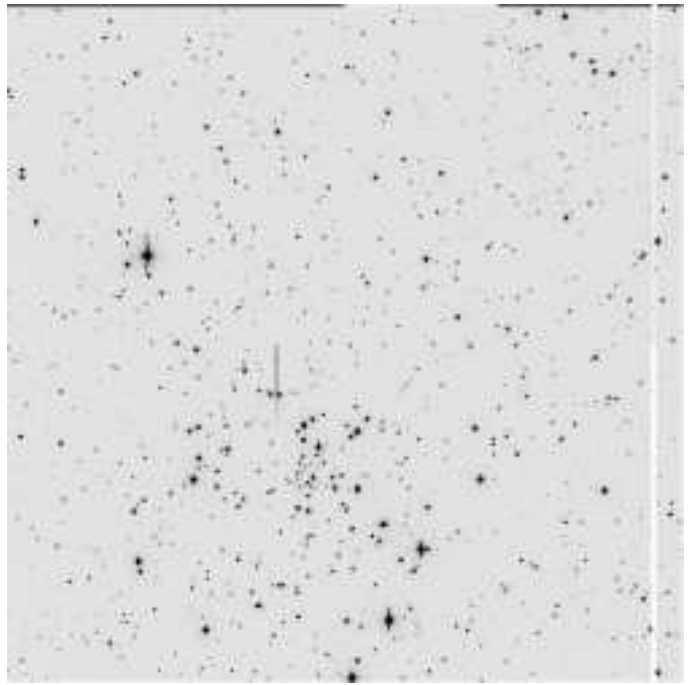}
   \caption{I = 600 sec image of NGC~2509. The field
is 13.5 arcmin on a side. North  is up, East to the left.}%
    \end{figure}

   \begin{figure}
   \centering
   \includegraphics[width=\columnwidth]{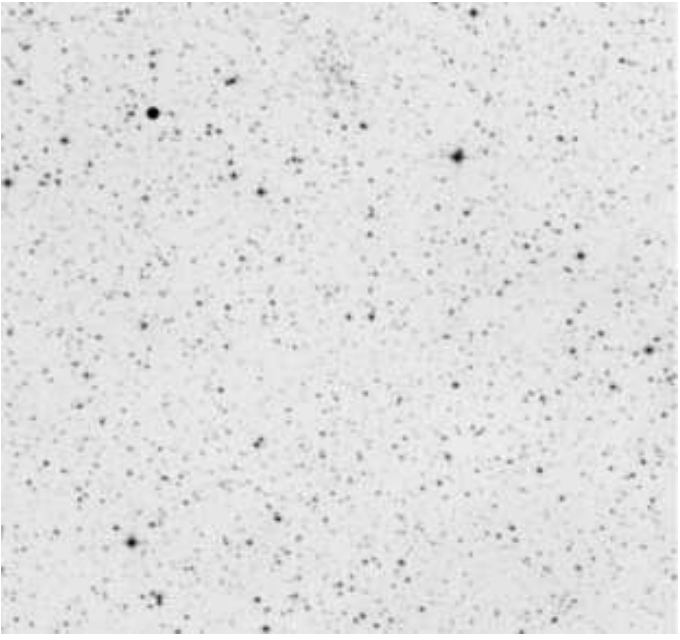}
   \caption{I = 900 sec image of VdB-Hagen~4. The field
is 20 arcmin on a side. North  is up, East to the left.}%
    \end{figure}

   \begin{figure}
   \centering
   \includegraphics[width=\columnwidth]{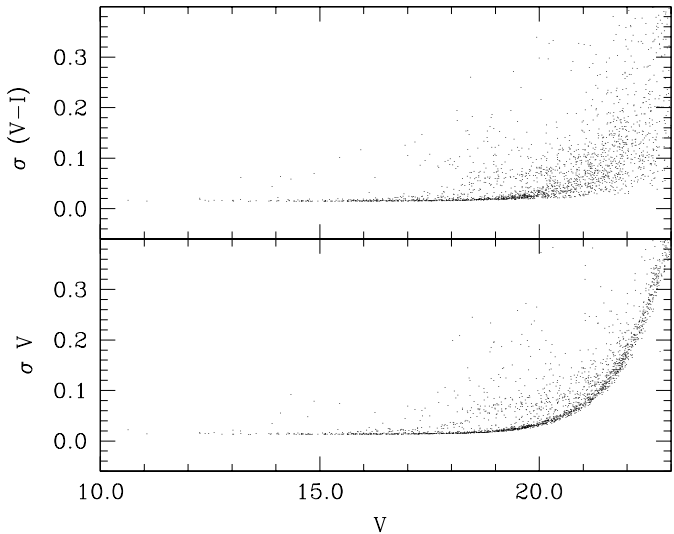}
   \caption{Photometric errors in V and (V-I) as a function of V mag.}%
    \end{figure}

\section{Observation and Data Reduction}

$V,I$ images centered on the star clusters
Collinder~74, Berkeley~27, Haffner~8, NGC~2509 and VdB-Hagen~4  were obtained with
the Cerro Tololo Inter-American Observatory  
0.9m and 1.0m telescopes, which are operated by the
SMARTS\footnote{http://www.astro.yale.edu/smarts/} consortium.

\subsection{0.9m observations}
Collinder~74, Berkeley~27, Haffner~8, NGC~2509
were observed with the 0.9m telescope on the nights
of December 3 and 4, 2005. The pixel scale of the 2048 $\times$
2046 Tek2k$\#$ CCD is 0$^{\prime\prime}$.369, yielding a field
of 13.5 $\times$ 13.5 arcmin on the sky. Both nights
were photometric with average seeing of 1.1 arcsec.
Due to the presence of very bright stars, Berkeley~27 and
NGC~2509 were not placed in the center of the CCD (see Figs. 2 and 4).\\
To secure a photometric calibration, the Landolt (1992)
standard fields SA~98, PG0231, TPhenix, G97 and G102
were observed at different airmasses several times
over the nights.
The photometric solutions of the two nights were
identical within the errors, so we derived a single
photometric solution for the two nights, using
the whole set of 170 standard stars.
The calibration equations have the form:\\
 
\noindent
$ v = V + (2.001\pm0.003) + (0.16\pm0.01) \times X + (0.007\pm0.002) \times (V-I)$ \\
$ i = I + (2.898\pm0.007) + (0.08\pm0.01) \times X + (0.031\pm0.004) \times (V-I)$ ,\\
 
\noindent
and the final {\it r.m.s} of the fitting 
turned out to be 0.020 and 0.022 for the V and I the pass-bands, respectively.\\

\subsection{1.0m observation}
The star cluster Vdb-Hagen~4 was observed with the 1.0m
telescope on the night of December 2, 2005.
The telescope is equipped with a new
4k$\times$4k CCD camera having a pixel scale of
0$^{\prime\prime}$.289/pixel which allows one to cover a field of
$20^{\prime} \times 20^{\prime}$.  
Three Landolt (1992) areas (TPhoenix, Rubin~149, and
PG~0231+006) were also observed to transform  the
instrumental magnitudes into the standard system.  The night was
photometric with an average seeing of 1.1 arcsec. 
In this case, the photometric solution for a grand-total of 
69 stars reads:\\

\noindent
$ v = V + (1.934\pm0.004) + (0.16\pm0.01) \times X + (0.014\pm0.005) \times (V-I)$ \\
$ i = I + (2.786\pm0.008) + (0.08\pm0.01) \times X - (0.012\pm0.008) \times (V-I)$ ,\\
 
\noindent
and the final {\it r.m.s} of the fitting
turns out to be 0.025 and 0.030 for the V and I the pass-bands,
respectively.
Due to the presence of a very bright star, the center of the frame
was chosen about 10 arcmin southward of the cluster.

\begin{table}
\centering
\caption{Log of photometric observations on December 2, 3 and 4, 2005.}
\begin{tabular}{lrrr}
\hline
Cluster& Date & Filter & Exp time (sec) \\
\hline
VdB-Hagen~4  & 02 December 2005 & V & 10, 180, 900\\
             &                  & I & 5, 180, 900\\
Collinder~74 & 03 December 2005 & V & 10, 20, 600\\
             &                  & I & 5, 2x10, 600\\
Haffner~8    & 03 December 2005 & V & 10, 20, 600\\
             &                  & I & 5, 10, 600 \\
NGC~2509     & 04 December 2005 & V & 10, 20, 600\\
             &                  & I & 5, 10, 600\\
Berkeley~27  & 04 December 2005 & V & 10, 20, 600\\
             &                  & I & 5, 10, 600\\
\hline
\end{tabular}
\end{table}

\noindent
The covered
areas are shown in Figs. 1 to 5, while Table~2  contains the
observational log-book.\\

\noindent
The data was reduced using the IRAF\footnote{IRAF is distributed by NOAO, which are
operated by AURA under cooperative agreement with the NSF.}
packages CCDRED, DAOPHOT, and PHOTCAL. Photometry was done following
the point spread function (PSF) method (Stetson 1987). The run
of photometric errors deriving form the PSF fitting is shown in Fig.~6\\
Global photometric errors have been estimated using the scheme
developed by Patat \& Carraro (2001), which take into account the
errors coming from the PSF fitting procedure (ALLSTAR), and the calibration
errors (corresponding to the zero points, color terms and extinction).
It turns out that stars brighter than
$V \approx 20$ mag turn out to have
global photometric errors (DAOPHOT internal plus calibration errors)
lower  than 0.25~mag in magnitude and lower than 0.35~mag in colour. \\
The final catalog consists of 2198, 4632, 4746, 4835 and
8353 point sources for Collinder~74, Berkeley~27, Haffner~8,
NGC~2509 and VdB-Hagen~4, respectively.\\
The photometry was finally aperture-corrected filter by filter
by using a suitable number (typically 15-20) of bright stars in the field.
The correction varied between 0.080 and 0.210 mag.\\

\noindent
Finally, the completeness corrections were determined by artificial-star
experiments on our data. Basically,
we created several artificial images by adding 
artificial stars to the original images. 
About 1000-5000 - depending on the frame -
were added to the original images.
In order to avoid the creation of overcrowding, in each experiment we added
at random positions only 15$\%$ of the original number of stars. The artificial
stars had the same color and luminosity distribution of the original sample.
In this way we found that the completeness level is better than  50$\%$ down to
V = 20.5.

\section{Clusters size}
We used star counts to estimate the size of each cluster
under study. The center of each cluster is searched for by inspecting
Figs. 1 to 5. In the case of VdB-Hagen~4 and Berkeley~27, the clusters
are offset with respect to the center of the frame, due to the presence
of very bright stars, which would not allow to observe the fainter
stars of the cluster.
Star counts are performed by counting the number of stars in concentric
ring 0.5 arcmin wide around the cluster center, and dividing by the 
ring surface. In the case of VdB-Hagen~4 we employed rings
0.25 arcmin wide.\\
To minimize absorption effects we have employed the I band,
the same we used in the series of Figs 1 to 5.\\
The aim of this procedure is to try to pick up as many cluster members
as possible limiting the contamination of field stars.
We consider as an estimate of cluster radius the distance from the center
at which the star counts keep flat, thus meaning that the level
of the field has been reached.\\
The results are shown in the various panels of Fig.~7. Examining this
figure, we can conclude that:

\begin{enumerate}
\item all the clusters are well confined within the observed
      region;
\item Collinder~74 is the cluster with the larger radius, amounting to
      4.0 arcmin; on the other hand, it  is the closest cluster, too;
\item all the other clusters are smaller, with the radii
      in the range between 1.0 (VdB-Hagen~4) to 3 (Berkeley~27) arcmin
\end{enumerate}

\noindent
These estimates nicely compare with Dias et al. (2001) values
of the clusters' diameters, which are based on visual inspection.
The values of the radius are reported in Table~3.\\
Having derived an estimate of the cluster radius, we can proceed
with realizations of each cluster population and nearby Galactic
fields.

   \begin{figure}
   \centering
   \includegraphics[width=\columnwidth]{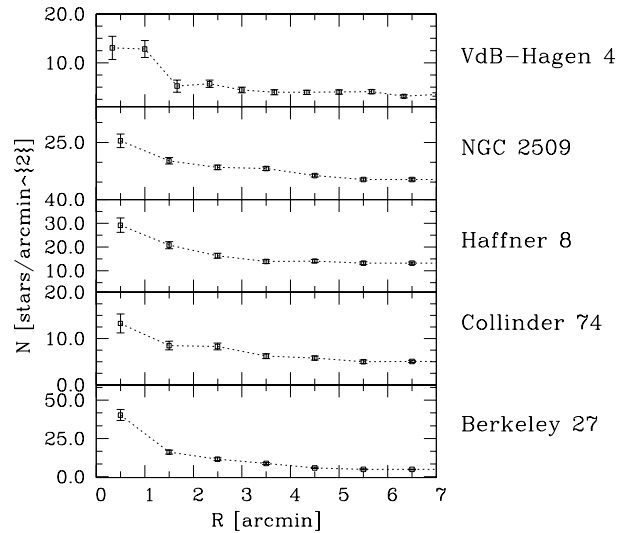}
   \caption{Estimating the clusters' radius through
    star counts}%
    \end{figure}

\section{Field star decontamination}
Open star clusters in the Galactic disk are known to suffer
from field star contamination which depending on the cluster
location inside the disk can make it very difficult
to derive the cluster fundamental parameters and in some
cases even to assess the cluster reality. \\
Here we make use of the results of the previous Sect. to obtain
a realization of the field star population in the direction of each
cluster, and then to statistically clean the cluster Color
Magnitude Diagram (CMD). This way we can better assess the reality
of the cluster, and, in positive cases, proceed to determine
its fundamental parameters. 
Star counts performed in the previous section are already suggestive 
of the reality of the clusters
under investigation. However differential reddening plays a role
in producing artificial clumps of stars, which look
like real clusters. The star distribution in the CMD
should help to get a clearer view of what is going on in the
Galactic directions we are looking at in this paper.\\

\noindent
In the series of Figs.~8 to 12, for each cluster we show the CMD of the stars enclosed
in a circular region having the radius derived in Section~3
(left panels), the CMD of an off-set field outside the cluster
region (middle panels), and the CMD of the cluster after having
statistically removed the field stars (right panels). We stress
the fact that the off-set fields have the same area of the cluster
fields. Besides, in the particular cases of Berkeley~27 and
VdB-Hagen~4 we applied the technique using more than one field
realization.\\
The statistical technique we employ works
as follows. We pick up an offset field star and look in the cluster
CMD for the closest star in color and magnitude, and remove this star
from the cluster star list. This procedure is repeated for all the 
offset field stars.\\
As amply discussed in the literature (see for instance Gallart et
  al. 2003, and Bertelli et
al. 2003), from which our code has been designed, the method depends on two crucial
parameters: the size of the searching ellipse, and the completeness
of the photometry. The completeness is known (see Sect.~2),
although  in this particular case is not a big issue, since
cluster and offset field come from the same frame and therefore
suffer from the same amount of incompleteness.
The size of the searching ellipse (see Gallart et al. 2003)
has to be fixed case by case by running many trials.

\subsection{Collinder~74}
This appears a sparse cluster (see Fig.~1) defined by a handful of bright
stars (Collinder 1933). Its low Galactic latitude is remarkable, and would mean
the cluster is nearby, if real. 
The results of the cleaning procedure is shown in Fig~.8.
With respect to the equal area field in the middle
panel, the cluster realization in the left panel shows a narrower Main
Sequence (MS), and the handful of bright stars we mentioned above.
The clean CMD in the right panel shows a MS which extends down to V$\sim$19.5.
However, it might be difficult
to interpret the distribution of the bright red stars and to
clearly define the cluster Turn Off Point (TO).

   \begin{figure}
   \centering
   \includegraphics[width=\columnwidth]{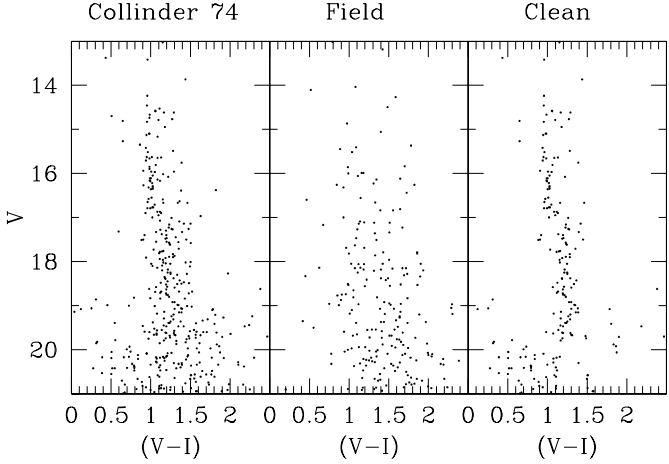}
   \caption{Field star decontamination: Collinder~74}%
    \end{figure}

   \begin{figure}
   \centering
   \includegraphics[width=\columnwidth]{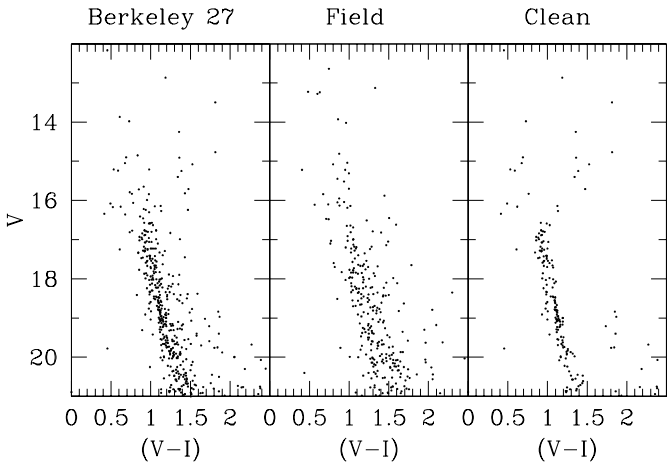}
   \caption{Field star decontamination: Berkeley~27}%
    \end{figure}

    \begin{figure}
   \centering
   \includegraphics[width=\columnwidth]{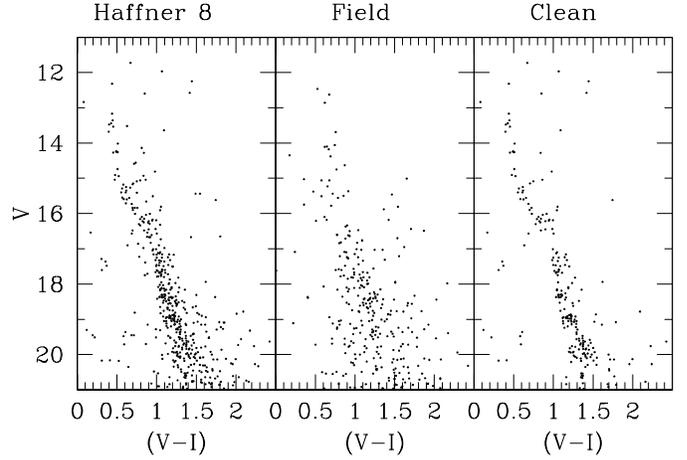}
   \caption{Field star decontamination: Haffner~8}%
    \end{figure}

   \begin{figure}
   \centering
   \includegraphics[width=\columnwidth]{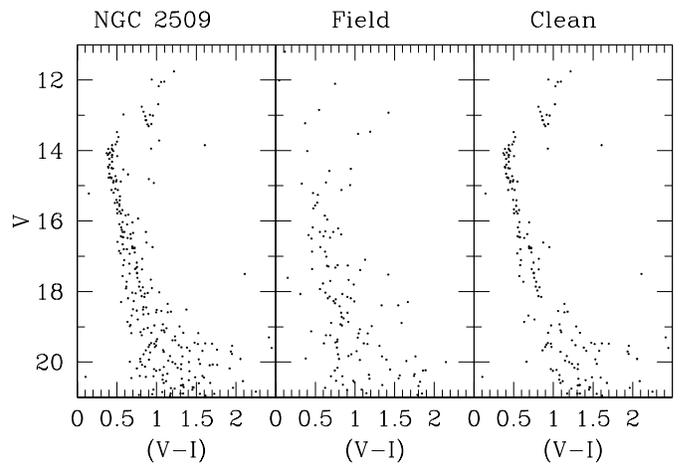}
   \caption{Field star decontamination: NGC~2509}%
    \end{figure}

\subsection{Berkeley~27}
This looks like a faint  cluster (see Fig.~2).
The CMD in the left panel of Fig.~9 exhibits
a MS more populated than in the equal area field realization in the middle panel.
The clean CMD shows a clear MS extending down to V$\sim$20.5, with a TO
located a V$\sim$17.0. The bunch of bright red ($(V-I) \geq 1.5$) 
stars visible in the left panel are very  probably cluster stars.
We conclude that Berkeley~27 is undoubtedly a real cluster.

\subsection{Haffner~8}
Haffner~8 is a sparse open cluster very similar in structure to
Collinder~74 (see Fig.~3). However the clean CMD (see Fig.~10, right panel)
reveals a much younger cluster, with only a few evolved stars.
The shape of the MS if a typical one for star clusters a few hundreds
million years old.\\
The TO points is located at V$\sim$13.5 and the MS is well populated down
to the limiting magnitude.

\subsection{NGC~2509}
The cluster is small but prominent (see Fig~4), and seems to harbor
a significant amount of bright stars.  The clean CMD in Fig.~11 (right panel)
shows a remarkable clump of bright stars typical of a star cluster
of the Hyades age (about 600 Myr). The shape of the clump closely resembles
that of clusters like NGC~2635 (Moitinho et al. 2006b), 
an intermediate age
open cluster as well located in the Third Galactic Quadrant.
The TO region is also very well
defined, with the TO at V$\sim$14.0. The color of the TO indicates
that the cluster suffers from a small amount of reddening, which is compatible
with its high Galactic latitude (see Table~1). The MS seems
to terminate abruptly at V$\sim$18.5, which again might indicate
that some low mass stars evaporation already occurred.

  \begin{figure}
   \centering
   \includegraphics[width=\columnwidth]{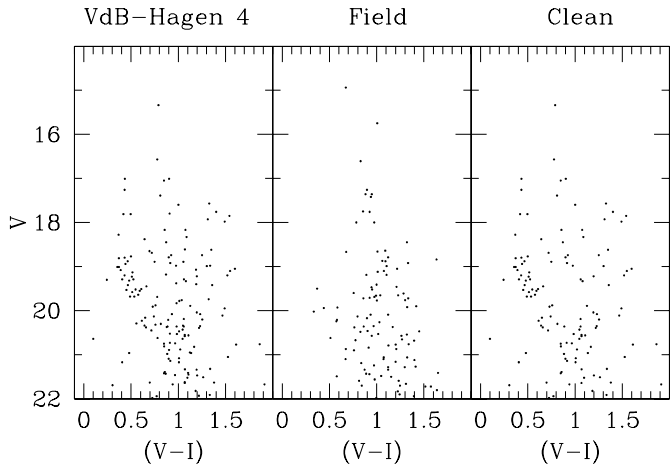}
   \caption{Field star decontamination: VdB-Hagen~4}%
    \end{figure}

\subsection{VdB-Hagen~4}
This cluster was never studied before, and was firstly detected by 
van den Bergh \& Hagen (1975). They describe the cluster as a poor ensemble
of stars, but as a  real feature both in the blue and in the red plates.
This is confirmed by inspecting Fig.~5, where we have placed the cluster 
in the upper part of the frame due to the presence of a disturbing very bright
star. Start counts confirm  the cluster is small. The CMD of the cluster (left
panel of Fig.~12) and the field (middle panel) look similar except for a tight
blue sequence in the cluster area, which witnesses the presence of a young
cluster/association. Indeed, the clean CMD maintains this sequence weakening
the field star contamination. The cluster MS is defined down to V$\sim$21
where it intersects the field MS. VdB-Hagen~4 is much probably a faint distant
star cluster. The contamination of field stars is really severe in this case,
and it is not possible to firmly probe the presence of evolved stars.

  \begin{figure}
   \centering
   \includegraphics[width=\columnwidth]{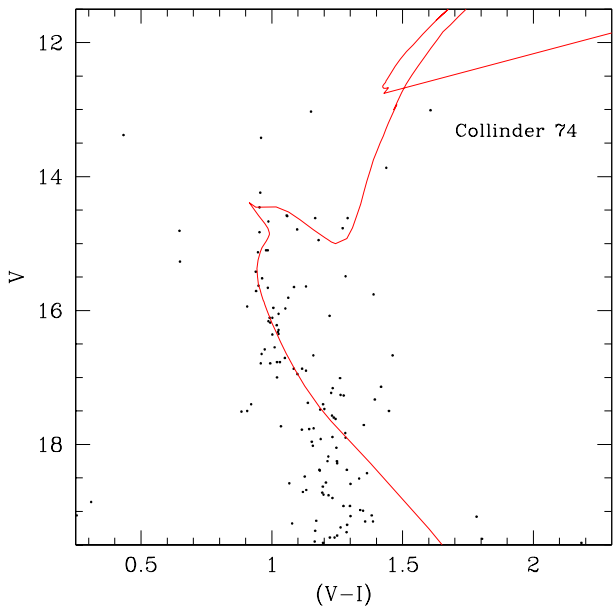}
   \caption{Isochrone solution for Collinder~74. The 3 Gyr isochrone
    is shifted by E(V-I)= 0.35 and V-M$_V$ = 11.75}%
    \end{figure}

  \begin{figure}
   \centering
   \includegraphics[width=\columnwidth]{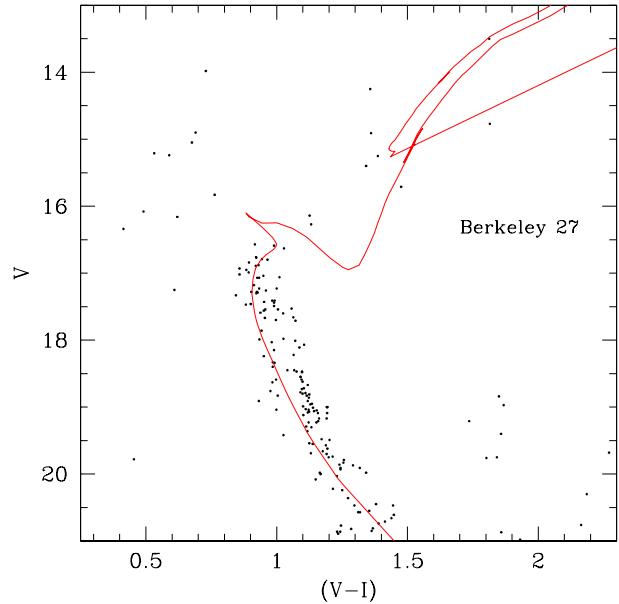}
   \caption{Isochrone solution for Berkeley~27. The 2 Gyr isochrone
    is shifted by E(V-I)= 0.35 and V-M$_V$ = 14.30}%
    \end{figure}

\section{Cluster Fundamental Parameters}
We now use the results of previous Section to infer estimates
of the cluster fundamental parameters.  For this purpose
we shall compare the clean CMDs with stellar isochrones
from Girardi et al. (2000). Lacking any spectroscopic
estimate of the metal abundance we shall adopt the conservative
solar metal abundance Z = 0.019. The results are summarized
in Table~3, where the radius, reddening and distance modulus
are reported together with their uncertainties. These reflect
the degree of freedom we have in displacing an isochrone still
keeping the goodness of the fit. In addition, in the same Table
the heliocentric and Galactocentric distance are listed
together with the best fit age.\\
\noindent
Moreover for each cluster we are going to recall any previous results
and make a detailed comparison

\subsection{Collinder~74}
This object was studied by Ann et al. (1999), who provide UBVI
photometry of a $11^{\prime}.5 \times 11^{\prime}.5$ field centered on
the cluster.
We compared our photometry with their one, and found that
$\Delta V = -0.072\pm0.049$, $\Delta (V-I) = -0.035\pm0.042$ (in the
sense their study minus the present one) from 135
stars in common brighter than V = 18.0, 
and concluded that the two studies are largely consistent.\\
These authors suggest that Collinder~74 is a star cluster 1.3 billion
years old located at 1.5 kpc from the Sun.\\
The clean CMD is shown again in Fig~13. Here a solar metallicity isochrone
is superimposed to the candidate cluster members.
After playing with several different ages, we come out with a best
fit age of 3 Gyr, which is shifted by E(V-I) = 0.35 and (V-M$_V$) = 11.75.
By adopting a normal extinction law, where
$\frac{E(V-I)}{E(B-V)}=1.245$, E(V-I)=0.35 turns into E(B-V) = 0.28,
which implies a heliocentric distance of 1.5 kpc.\\
The reason for which we adopted this fit resides in the nice
reproduction of the TO region. As a consequence the bulk of bright
red stars mentioned above are interpreted as sub-giant and giant branch stars.
A couple of brighter and redder stars cannot be fitted, and might mean
that the cluster is metal richer, in case they were members.
The MS seems to abruptly drop at V$\sim$18, which is compatible
with the cluster having lost most of its low mass stars.\\
As guessed in the previous section, the cluster is located close to the Sun.
Its position and height above the Galactic plane (Z =  270 pc)
are reasonable for a cluster of this age.
We believe that the significantly older age we found we respect
to previous investigation is due to the better treatment of the
field star contamination, which allows the various features of the CMD
to stand out  more clearly. The lack of any Red Giant Branch (RGB) clump
is not unusual for a cluster of this age, since several other old
clusters are known to have a few clump stars, if any (Phelps et al. 1994).

\subsection{Berkeley~27}
Berkeley~27  (Setteducati \& Weaver 1962) is also known as
Biurakan~11.\\
The only previous investigation we are aware of is by Hasegawa et al. (1999).
They obtained BVI photometry with of a $5^{\prime}.1 \times 5^{\prime}.1$ 
field 
centered on the cluster. Their analysis leads to these values
for the cluster parameters: an age of 2.0 billion years, a distance of 13.5
kpc from the Galactic center and a reddening E(V-I) = 0.30 .
The result of our fit is shown in Fig.~14.
We basically find the same overall solution proposed by Hasegawa et al (1999).
We confirm that the probable age is around 2.0 Gyr, and the corresponding
isochrone has to be shifted by E(V-I) = 0.35 and (V-M$_V$)=14.30.
This in turn yields a heliocentric distance of 4.8 kpc, and a height
above the Galactic plane Z $\sim$ 250 pc.
This cluster shares with Collinder~74 the lack a clear RGB and clump, but
exhibits a wider MS, probably due some uncorrected field star contamination,
a probable percentage of unresolved binaries and some differential reddening.
This latter might in fact plays a more important role than in the case
of Collinder~74, due to lower Galactic latitude of Berkeley~27.

\subsection{Haffner~8}
For this cluster only the old photographic study by Fenkart et al. (1972) is available.
These authors suggest the cluster is 500 million year old, and it is located
at 1.7 kpc from the Sun. Strangely, they assign to the cluster zero absorption,
which one would not expect judging from the location of the cluster relatively
low onto the Galactic plane. In fact, for instance, from Schlegel et al. (1998)
maps one can infer E(B-V) = 0.66 which would imply A$_V$ = 2.0 at infinity,
by assuming the normal extinction law $R_V = \frac{A_V}{E(B-V} = 3.1$.
Zero reddening would mean the cluster lies very close to the Sun.
However, most probably something wrong occurred with the photographic
photometry. Unfortunately, we could only compare the V filter with
them, and found that $\Delta V = -0.16\pm0.21$ for 32 stars in common
brighter than V = 15.\\
Our clean CMD is presented in Fig.~15. The 500 Myr isochrone which provides
the best fit is shifted by E(V-I) = 0.30 and V-M$_V$ = 12.80.
By assuming the normal extinction law we get  E(B-V)=0.24, which is reasonable
for a cluster placed relatively low onto the Galactic plane. 
We remind that the reddening at infinity in this direction is E(B-V)=0.66 (Schlegel et al. 1998).\\
The fit is very good, both in the TO region and along the MS.
Also the evolved part of the CMD is nicely reproduced. The derived parameters
for Haffner~8 are listed in Table~3.

\subsection{NGC~2509}
This clusters has been studied several times in the last few years,
but different analysis led to very discrepant results. While
Sujatha \& Babu (2003) found that the cluster is very old (8 Gyrs),
Ahumada (2000) concluded that NGC~2509 is only 1 Gyr old and has
E(B-V)=0.22.
The recent study by Tadross (2005) based on 2MASS
does not solve the problem
since it is not clear if the author assigns to the cluster the age of 1.6
or 6.1 Gyr. Moreover the same author assigns to the cluster
a suspicious value of E(B-V)=0.0\\
Even the reddening
found by Ahumada does not seem correct, since
the Schlegel et al. (1998) maps predict in this
direction an absorption of E(B-V) = 0.16. Since this is the
reddening value at infinity, NGC~2509 should be less reddened.
However, the visual inspection of the cluster CMD on the overall
 favors Ahumada (2000) results. 
In fact, the photometry by Sujatha \& Babu (2003) is highly
suspicious, as already found in the recent study of NGC~2401 (which they observed
the same night as NGC~2509) by Baume et al. (2006).\\
Our isochrone solution is illustrated in Fig.~16.
Here the 1.2 Gyrs isochrone for solar metallicity has been shifted
by E(V-I) = 0.08 and V-M$_V$ = 12.50. While the TO region and
the whole MS is nicely reproduced, this isochrone fails
to reproduce the color of the RGB and clump. 
This may be due to the fact that the exact metallicity is not solar,
an issue which can be solved only by a spectroscopic
study of the clump stars.\\
On the other hand the mixing length parameter is an important 
ingredient of stellar evolutionary models which play a role in determining
the precise color of RGB stars. For instance the discussion
in Palmieri et al. (2002) clearly demands a revision of
the (V-I) versus temperature relation, 
which may help to solve
the discrepancy between theory and observations
also in a metallicity regime higher than that of globular clusters.

  \begin{figure}
   \centering
   \includegraphics[width=\columnwidth]{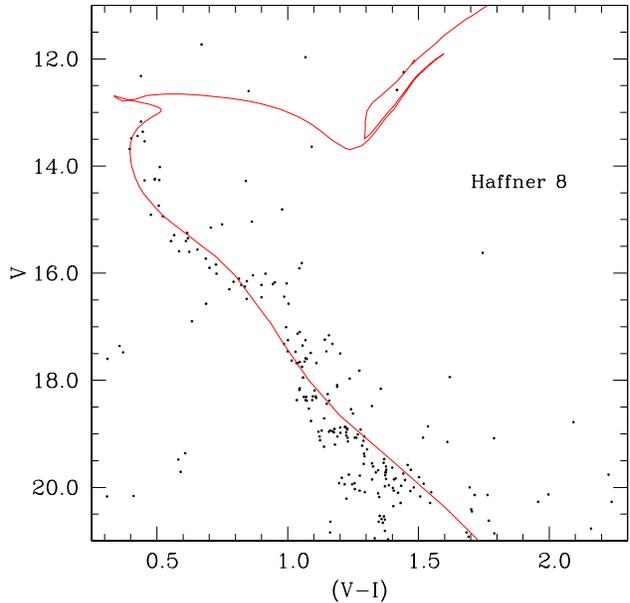}
   \caption{Isochrone solution for Haffner~8. The 0.5 Gyr isochrone
    is shifted by E(V-I)= 0.30 and V-M$_V$ = 12.80}%
    \end{figure}

 \begin{figure}
   \centering
   \includegraphics[width=\columnwidth]{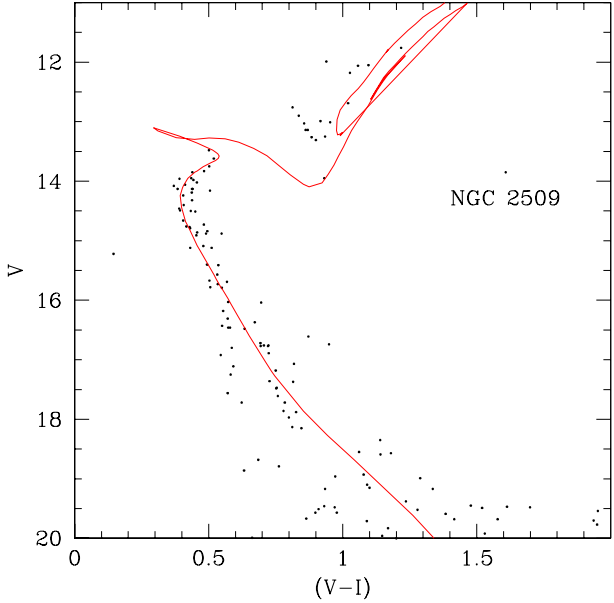}
   \caption{Isochrone solution for NGC~2509. The 1 Gyr isochrone
    is shifted by E(V-I)= 0.08 and V-M$_V$ = 12.50}%
    \end{figure}

  \begin{figure}
   \centering
   \includegraphics[width=\columnwidth]{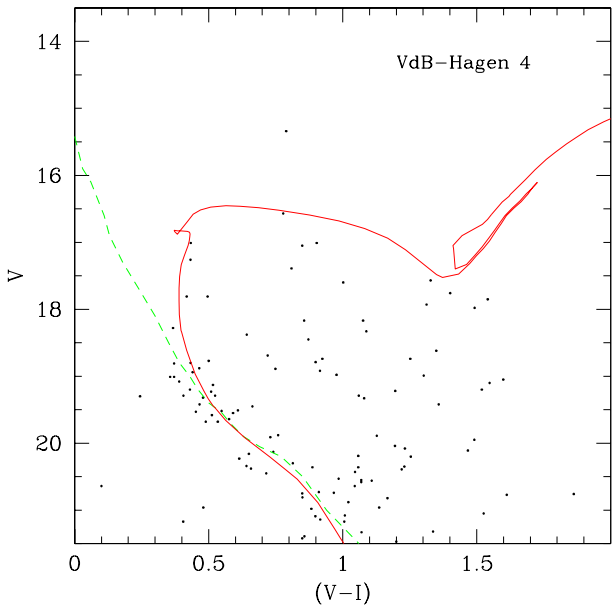}
   \caption{Isochrone solution for VdB-Hagen~04. The 0.2 Gyr isochrone
    is  shifted by E(V-I)= 0.45 and V-M$_V$ = 17.50}%
    \end{figure}

\begin{table*}
\caption{Fundamental parameters as derived from isochrone fitting}
\begin{tabular}{ccccccccc}
\hline
\hline
\multicolumn{1}{c}{Name} &
\multicolumn{1}{c}{Apparent ~Radius}  &
\multicolumn{1}{c}{Absolute ~Diameter} &
\multicolumn{1}{c}{$E(V-I)$}  &
\multicolumn{1}{c}{$(V-M_V)$} &
\multicolumn{1}{c}{$d_{\odot}$} &
\multicolumn{1}{c}{$Z$} &
\multicolumn{1}{c}{$d_{GC}$}&
\multicolumn{1}{c}{$Age$} \\
\hline
& arcmin & pc & mag & mag & kpc & pc & kpc & Myr \\
\hline
Collinder 74    & 4.0$\pm$0.5 &  3.5 & 0.35$\pm$0.05 & 11.75$\pm$0.10 &  1.50 & -270 &  9.90 & 3000\\
Berkeley  27    & 3.0$\pm$0.5 &  4.2 & 0.35$\pm$0.05 & 14.30$\pm$0.10 &  4.80 &  200 & 12.90 & 2000\\
Haffner 8       & 2.5$\pm$0.5 &  3.8 & 0.30$\pm$0.05 & 12.80$\pm$0.10 &  2.60 &   60 & 10.40 &  500\\
NGC 2509        & 2.5$\pm$0.5 &  4.2 & 0.08$\pm$0.05 & 12.50$\pm$0.10 &  2.90 &  200 & 10.30 & 1200\\
VdB-Hagen 4     & 1.0$\pm$0.5 & 11.2 & 0.45$\pm$0.05 & 17.50$\pm$0.20 & 19.30 & 2300 & 23.50 & $\sim$ 200\\
\hline\hline
\end{tabular}
\end{table*}

\subsection{VdB-Hagen~4}
In Fig.~17 we show the CMD of VdB-Hagen~4 corrected for field
star contamination. We superpose an isochrone of 200 Myr shifted
by E(V-I) = 0.43 ( E(B-V) = 0.32)  and (V-M$_V$)= 17.50. As for the reddening,
the value at infinity derived from Schlegel et al. (1998) maps is
E(B-V) = 0.49.
The fit here is mainly based on
the shape of the MS, and looks reasonable.  This is confirmed
by the empirical Zero Age Main Sequence (ZAMS, dashed line) from Schmidt-Kaler (1982) which is displayed
is the same plot and which has been shifted by the same apparent
distance modulus and reddening.\\
This fit demonstrates that the star clustering VdB-Hagen~4 is really extreme:
it is a group of young stars located at about  19 kpc from the Sun (see Table~3).
This distance and the Galactic coordinates of the group are compatible
with it being part of the outer (Norma-Cygnus) spiral arm (see Moitinho et al.
2006a). The height Z  above the Galactic plane results about 2 kpc, which
is compatible with the maximum warping of the Galactic disk in the Third
Galactic Quadrant (Carraro et al. 2006, V\'azquez et al. 2006).

\section{Conclusions}
We have used new CCD photometry in the V and I pass-bands to study
the stellar content of 5 Galactic fields containing the star
clusters Collinder~74, Berkeley~27, Haffner~8, NGC~2509 and
VdB-Hagen~4. The basic parameters we estimated
are summarized together with their uncertainties in Table~3.\\
\noindent
The most interesting results we found can be summarized as follows:

   \begin{description}
      \item $\bullet$ Collinder~74 is found to be significantly older than previous
             studies. With an age of 3 Gyr, if falls in an age interval where
             only a few old open clusters are known;
             therefore this cluster contributes to make less important the claim
             that a burst of star formation occurred around 5 Gyrs ago in the Galactic
             disk (Carraro et al. 2005);
      \item $\bullet$ we clarify some discrepancies present in the literature for NGC~2509
             and Haffner~8; the claim that NGC~2509 is very old and possesses zero
             reddening is found to have no observational support;
             the position of the RGB and clump of NGC~2509 poses an interesting
             challenge for theoretical models, and more work is needed for
             this cluster, especially a spectroscopic  estimate of the metal content;
      \item $\bullet$  the Haffner~8 zero reddening found by previous investigations
             is shown to be most probably due 
             to problems in the calibration of the photographic photometry;
      \item $\bullet$ as for Berkeley~27, this is the only cluster for which we find nice
             agreement with previous investigation ;
      \item $\bullet$ VdB-Hagen~4  is an extremely interesting small cluster;
            according to our analysis it is a young cluster belonging to the outer Galactic
            spiral arm in the extreme periphery of the Galactic disk. To better elucidate
            the stellar populations in this direction, UBV photometry
            is needed;
      \item $\bullet$ the cluster radii, expressed in parsec, are
             well within  the typical radius of open star clusters
             in the Galactic disk (Janes et al. 1988). The only
             exception is VdB-Hagen~4, which has a significantly
             larger radius. This is however compatible with the trend
             of open cluster size as a function of the height from the
             Galactic plane (Chen et al. 2004).
   
   \end{description}

\noindent
With the exception of Haffner~8, the clusters of this sample exhibit
heights above the Galactic plane larger than the
typical Population I thin disk scale height ($\approx 70$ pc). 
We argue that this is probably an age effect, and in fact old open
clusters (say older than the Hyades) do show on the average large
heights above the Galactic plane (Friel 1995).

\begin{acknowledgements}
GC work is supported by a grant of the Departmento de Astron\'omia
de Universidad de Chile. EC acknowledges FONDAP No. 15010003.
\end{acknowledgements}


\begin{thebibliography}{}

   \bibitem[2000]{ahumada} Ahumada, J.A., ASPC, 198, 43

   \bibitem[1999]{ann} Ann, H.B., Lee, M.G., Chun, M.Y., et al., 1999,
    JKAS, 32, 7

  \bibitem[2006]{bau} Baume, G., Moitinho, A., V\'azquez, R.A.,
  Solivella, G., Carraro, G., Villanova, S., 2006, MNRAS 367, 1441

  \bibitem[2003]{ber} Bertelli, G., Nasi, E., Girardi, L., Chiosi, C.,
  Zoccali, M., Gallart, C., 2003, AJ 125, 770

   \bibitem[1975]{bergh} van den Bergh, S., Hagen, G.L., 1975, AJ, 80, 11


   \bibitem[2005]{carraro} Carraro, G., Geisler, D., Moitinho, A., Baume, G., V\'azquez, R.A.,
    2005, A\&A, 442, 197


   \bibitem[2006]{carraro} Carraro, G., Moitinho, A., Zoccali. M., V\'azquez, R.A.,
    Baume, G., 2006, AJ, in press

   \bibitem[2004]{chen} Chen, W.P., Chen, C.W., Shu, C.G., 2004, AJ 128, 2306


   \bibitem[1931]{colli} Collinder, O., 1931, Ld.An., 2

   \bibitem[1972]{fenkart} Fenkart, R.P., Buser, R., Ritter, H., et al.,    1972, A\&AS, 7, 487

   \bibitem[1995]{friel} Friel, E.D., 1995, ARA\&A 33, 381


   \bibitem[2003]{galla}Gallart, C.,  Zoccali, M.,  Bertelli, G.,
	Chiosi, C.,  Demarque, P.,  Girardi, L.,  Nasi, E., Woo,
	J.-H., Yi, S., 2003, AJ 125, 742	

   \bibitem[2000]{girardi} Girardi, L., Bressan, A., Bertelli, G., 
       Chiosi, C., 2000, A\&AS, 114, 371
   
   \bibitem[1988]{janes} Janes, K.A., Tilley,  C., Lynga, G., 1988, AJ
   95, 771

   \bibitem[1992]{landolt} Landolt, A.U., 1992, AJ, 104, 372

   \bibitem[2006a]{moitinho} Moitinho, A., V\'azquez, R.A., Carraro, G.,
    Baume, G., Giorgi, E.E, \& Lyra, W. 2006a,
      MNRAS, 368, L77

   \bibitem[2006b]{moitinho} Moitinho, A., Carraro, G., Baume. G.,
     V\'azquez, R.A., 2006, A\&A, 445, 939

   \bibitem[2006]{momany} Momany, Y., Zaggia, S., Gilmore, G., Piotto, G.,
    Carraro, G., Bedin, L., \& De Angeli, F.,
      2006, A\&A, 451, 515

   \bibitem[2002]{newberg} Newberg, H.J., Yanny, B., Rockosi, C.,
    et al., 2002, ApJ, 569, 245 

    \bibitem[2002]{palmieri} Palmieri, R., Piotto, G., Saviane, I., Girardi, L.,
     Castellani, V., 2002, A\&A 292, 115

   \bibitem[2001]{pata} Patat, F., Carraro, G., 2001, MNRAS 325, 1591

   \bibitem[1994]{phelps} Phelps, R.L., Janes, K.A., Montgomery, K.A.,
    1994, AJ, 107 1079 

   \bibitem[2003]{sajatha} Sajatha, S., Babu, G.S.D., 2003, BASI, 31, 9

   \bibitem[1998]{schlegel} Schlegel, D.J., Finkbeiner, D.P., Davis, M.,
     1998, ApJ, 500, 525

    \bibitem[982]{sch82}
    Schmidt-Kaler, Th., 1982, Landolt-B\"ornstein, Numerical data and Funct
    ional Relationships in Science and Technology, New Series, Group VI, Vol. 2(b),
    K. Schaifers and H.H. Voigt Eds., Springer Verlag, Berlin, p.14

   \bibitem[2005]{tadross} Tadross, A.L., 2005, JKAS 38, 357

   \bibitem[1962]{setteducati} Setteducati, A.F., Weaver, M.F., 1962,
   Newly Found Stellar Clusters, published by Radio Astr. Lab. University
   of California, Berkeley


   \bibitem[1987]{stetson} Stetson, P.B., 1987, PASP, 99, 191

   \bibitem[2006]{vazquez} V\'azquez, R.A., May, J., Carraro,  G., Bronfman, L.,
   Moitinho, A., Baume, G., 2006, ApJ, submitted

\end{thebibliography}
\end{document}